\documentclass[12pt,preprint]{aastex}

\IfFileExists{srcltx.sty}{\usepackage[active]{srcltx}}

\usepackage{natbib}
\usepackage[colorlinks,urlcolor=blue,citecolor=blue,linkcolor=blue]{hyperref}

\usepackage{epsfig}
 \usepackage{verbatim}
 \usepackage{graphicx}

\shorttitle{WEAK REDSHIFT DEPENDENCE}
\shortauthors{ESSEY ANS KUSENKO}

\begin{document}

\title{On weak redshift dependence of gamma-ray spectra of distant blazars}

\author{Warren~Essey\altaffilmark{1} and Alexander Kusenko \altaffilmark{2,3}}

\altaffiltext{1}{International Center for Computer Science, University of California, Berkeley, CA 94720, USA }

\altaffiltext{2}{Department of Physics and Astronomy, University of
California, Los Angeles, CA 90095-1547, USA}

\altaffiltext{3}{Kavli Institute for the Physics and Mathematics of the Universe,
University of Tokyo, Kashiwa, Chiba 277-8568, Japan}


\begin{abstract}
Line-of-sight interactions of cosmic rays provide a natural explanation of the hard gamma-ray spectra of distant blazars, which are
believed to be capable of producing both gamma rays and cosmic rays.  For sources with redshifts $z\gtrsim 0.1$, secondary gamma rays produced 
in cosmic-ray interactions with background photons close to an observer can dominate over primary gamma  rays originating at the source. 
The transition from one component  to another is accompanied by a change in the spectral index depending on the source redshift.  
We present theoretical predictions and  show that they agree with the data from Fermi Large Area Telescope. This agreement, combined with the spectral data from Atmospheric Cherenkov Telescopes, provides evidence of cosmic ray acceleration by active galactic nuclei and opens new opportunities for studying photon backgrounds and intergalactic magnetic fields. 
\end{abstract}

\maketitle

\newpage

\section{Introduction}
Active galactic nuclei (AGN) are powerful sources of gamma rays, and they are widely believed to produce cosmic rays. 
It was recently proposed that the hardness of gamma-ray spectra of distant blazars can be naturally explained by the line-of-sight interactions of cosmic rays accelerated in the blazar jets~\citep{2010APh....33...81E,2010PhRvL.104n1102E,2011APh....35..135E,2011ApJ...731...51E}.  While primary gamma rays emitted by the blazar are attenuated in their interactions with extragalactic background light (EBL)~\citep{1998ApJ...493..547S}, cosmic rays with energies $10^{16} - 10^{19}$~eV can cross cosmological distances and can produce secondary gamma rays in their interactions with the background photons. The predicted spectra of these secondary gamma rays are very robust and are not sensitive to the uncertainties in the level of EBL or the spectrum of protons at the source, except for the cosmic-ray luminosity. The  predictions are in excellent agreement with the data~\citep{2010APh....33...81E,2010PhRvL.104n1102E,2011ApJ...731...51E,2011arXiv1107.5576M}.  
In the absence of cosmic-ray contribution, some unusually hard intrinsic spectra~\citep{2007ApJ...667L..29S,2011ApJ...740...64L,2011arXiv1110.3739D} or hypothetical new particles~\citep{2007PhRvD..76l1301D} have been invoked to explain the data.

The success of this picture lends support to the hypothesis of cosmic ray acceleration in AGNs.  Identifying the origin of ultrahigh-energy cosmic rays (UHECR) is difficult because the deflections of protons and ions in the galactic magnetic fields weaken the correlations of UHECR arrival directions with the positions of their sources~\citep{2007Sci...318..938T,2008APh....29..188P}. Furthermore, 
a contribution of transient galactic sources of high-energy nuclei can further complicate identification of extragalactic sources~\citep{2010PhRvL.105i1101C}.
In contrast, a definitive confirmation of the line-of-sight interactions~\citep{2010APh....33...81E,2010PhRvL.104n1102E,2011ApJ...731...51E,2011arXiv1107.5576M} would 
make possible gamma-ray astronomical observations of cosmic rays while they are still well outside the reach of strong galactic magnetic fields.  One can use the existing gamma-ray data to set lower limits on the power of cosmic ray acceleration in blazars~\citep{2012ApJ...745..196R}.

At small distances, primary gamma rays dominate the observed signals of blazars, and it is only at redshifts $z\gtrsim 0.15$ that the cosmic-ray induced
contribution comes to dominate because the primary gamma rays are attenuated by their interactions with EBL. The existence of two independent components implies a change in the spectral index and the existence of some intermediate range of redshifts in which one or the other component can be seen, depending on the individual properties of blazars. We will identify the spectral properties of both components, and we will use Second AGN catalog from Fermi LAT~\citep{2011arXiv1108.1420T} to test our predictions. 

For primary gamma rays, \cite{2006ApJ...652L...9S,2010ApJ...709L.124S} have derived a simple scaling law which explained the redshift dependence of spectra of the nearby blazars.  Although their original fit has some additional parameters, the spectral evolution due to absorption over distance $d\approx z/H_0$ can be described by a simplified expression:

\begin{equation}
F\propto \frac{ e^{-d/\lambda_\gamma}}{d^2} \ E^{-(\Gamma_s +DH_0 d )}
\end{equation}
Here $\lambda_\gamma$ is the distance at which EBL opacity to TeV gamma rays is  of the order of $1$, $\Gamma_s$ is the intrinsic spectral index of  gamma rays at the source, $H_0$ is the Hubble constant, and $D$ is a parameter that describes spectral change due to attenuation in gamma-ray interactions with EBL~\citep{2006ApJ...652L...9S,2010ApJ...709L.124S}.
This simple law, as well as its more precise implementations \citep{2006ApJ...652L...9S,2010ApJ...709L.124S}, provides an excellent fit to the data at small redshifts. However, at higher redshifts, there is a significant deviation from the \cite{2006ApJ...652L...9S,2010ApJ...709L.124S} relation:  the spectral index evolution with redshift is much slower, as shown in Figure~\ref{fig:fit}. 

We note that most of the low-redshift sources are high-synchrotron-peaked (HSP) blazars, while the distant sources are dominated by intermediate-synchrotron peaked (ISP) blazars and flat spectrum radio quasars (FSRQ). For ISP and FSRQ the GeV signal may be at or below the Compton peak and our analysis above does not take this spectral variation into account. However, this effect would increase $\Delta\Gamma$, because the variation implies some additional softening due to moving past the Compton peak, which is not supported by the data.  TeV spectra, if they are secondary gamma rays produced along the line of sight, do not depend significantly on the gamma-ray or proton spectra of their sources~\citep{2010APh....33...81E,2010PhRvL.104n1102E,2011ApJ...731...51E,2011arXiv1107.5576M,2012ApJ...745..196R}.  The dependence on the EBL 
model~\citep{2010ApJ...712..238F,2008A&A...487..837F,2006ApJ...648..774S,2009MNRAS.399.1694G,2011ApJ...733...77O} is very weak~\citep{2011ApJ...731...51E}. 
Thus, the spectral variation does not affect our conclusion that the behavior in Figure~\ref{fig:fit} is consistent with a new component taking over and dominating the signal for $z\gtrsim 0.15$.  
For the same reason, our best-fit line in Figure~\ref{fig:fit} does not depend on the choice of the EBL model.

Line-of-sight interactions of cosmic rays can account for the hard spectra of distant blazars because, in this case, the observed multi-TeV gamma rays are produced in interactions of cosmic rays with the background photons relatively close to Earth ~\citep{2010APh....33...81E,2010PhRvL.104n1102E,2011ApJ...731...51E,2011arXiv1107.5576M}. For this reason, the distance to the source is much less important 
than in the case of primary sources.  One, therefore, expects the spectra of secondary gamma rays to exhibit a slower change with redshift.

\section{Softening of a two-component spectrum}

We would like to generalize the \cite{2006ApJ...652L...9S,2010ApJ...709L.124S} scaling law to include the additional component at high redshift. 
The fluxes of primary gamma rays produced at the source and of secondary gamma rays produced in line-of-sight interactions of protons scale with distance $d$ as follows~\citep{2011ApJ...731...51E}:  

\begin{eqnarray}
F_{\rm primary,\gamma}(d)& \propto &  \frac{1}{d^2} e^{-d/\lambda_\gamma} \label{exponential} 
\end{eqnarray}

\begin{eqnarray}
F_{\rm secondary,\gamma}(d)& \propto &  \frac{\lambda_\gamma}{d^2}\Big(1-e^{-d/\lambda_\gamma}\Big)  \\
& \sim &  \left \{ 
\begin{array}{ll}
1/d, & {\rm for} \ d \ll \lambda_\gamma, \\ 
1/d^2, & {\rm for} \ d\gg \lambda_\gamma .
\end{array} \right.
\end{eqnarray}

Obviously, for a sufficiently distant source, secondary gamma rays must dominate because they do not suffer from exponential suppression as in Equation~(\ref{exponential}). The predicted spectrum of $\gamma$-rays turns out to be similar for all the distant AGN.  \cite{2010APh....33...81E,2010PhRvL.104n1102E,2011ApJ...731...51E} have calculated the spectra for redshifts of 3C279, 1ES~1101-232, 3C66A, 1ES0229+200, and several other blazars, all of which yield a remarkably good (one-parameter) fit to the data~\citep{2010APh....33...81E,2010PhRvL.104n1102E,2011ApJ...731...51E}.  

Based on our numerical results using a Monte Carlo propagation code described by \cite{2010APh....33...81E,2010PhRvL.104n1102E,2011ApJ...731...51E}, we find that the spectra have a weak redshift dependence and, in the TeV energy range, for $0.2 \lesssim z \lesssim 0.6$, it can be approximated by the following simple relation: 
\begin{equation}
\Gamma_{\rm TeV} \simeq \Gamma_p+\alpha z, 
\end{equation} 
where $\Gamma_p$ is a constant and $\alpha\approx 1$.

Let us now consider a flux of TeV gamma rays which is the sum of two components that have the above mentioned scaling with distance: 
\begin{eqnarray}
 F_{\rm TeV} & =& F_1 \,  \frac{1}{d^2}\exp(-d/\lambda_\gamma)  \ E^{-(\Gamma_s +DH_0 d)} + \nonumber \\
& & F_2\, \frac{1}{d^2}\Big(1-e^{-d/\lambda_\gamma}\Big) E^{-(\Gamma_p+\alpha H_0 d)} \\
& =& \frac{1 }{d^2} \, \left [ e^{-d/\lambda_\gamma} \, \left(
F_1 E^{-(\Gamma_s +DH_0 d)} - F_2  E^{-(\Gamma_p+\alpha H_0 d)} \right) \right. \nonumber \\ & & \left . 
+ F_2 \, E^{-(\Gamma_p+\alpha H_0 d) } \right ]\label{newscaling}
\end{eqnarray}

While the overall $1/d^2$ factor does not affect the spectral index, the exponential suppression of the first term in squared brackets in Equation(\ref{newscaling}) guarantees a sharp change from the 
\cite{2006ApJ...652L...9S,2010ApJ...709L.124S} scaling law to a flatter scaling law which shows only a weak redshift dependence.  The change occurs when the distance $d$ is of the order of $\lambda_\gamma$, i.e. at a distance from the source where EBL optical depth approaches 1.  Based on our numerical calculations, and in agreement with \cite{2006ApJ...652L...9S}, the corresponding redshift is $z\approx H_0 d \approx 0.1$.  Taking into account that $F_1\gg F_2$, one can write an approximate scaling law as

\begin{eqnarray}
z^2 \, F_{\rm TeV}\, & \propto & e^{-z/0.1}\, F_1 \ E^{-(\Gamma_s +D z)} + F_2 E^{-(\Gamma_p+\alpha z)}
\end{eqnarray}

At lower energies, in the GeV energy range, the flux is expected to show very little attenuation for $z\lesssim 0.5$ and to follow the simple relation
\begin{eqnarray}
z^2\, F_{\rm GeV} \, & \propto &  \tilde{F}_1 \ E^{-\Gamma_s } 
\label{eq:flux_two_components}
\end{eqnarray}

Thus we expect that $\Delta \Gamma = \Gamma_{\rm TeV} - \Gamma_{\rm GeV}$ should exhibit the following behavior: 
\begin{eqnarray}
\Delta \Gamma \simeq &  \left \{ 
\begin{array}{ll}
Dz & {\rm for} \ z\lesssim 0.1, \\ 
(\Gamma_p-\Gamma_s)+\alpha z, & {\rm for}  \ z\gtrsim 0.1 .
\end{array} \right.
\label{eq:deltaG}
\end{eqnarray}

For practical reasons, it is easier and more instructive to compare the spectral slopes given by Equation~(\ref{eq:deltaG}) with the data, 
rather than to fit the fluxes in Equation~(\ref{eq:flux_two_components}). 

To select distant sources that are likely to be powerful sources of cosmic rays (see Table~1), we applied two selection criteria: we selected gamma-ray emitters which (i) have been observed at energies where the optical depth for pair production $\tau $ greatly exceeds one,  and which (ii) showed no short-scale time variability at these relevant energies.
We emphasize that these sources can show variability at lower energies, where the energy dependent optical depth $\tau (E) \lesssim 1$. 
Time variability has been reported for integrated flux at $E>200$~GeV for 3C 
66A \citep{2011ApJ...726...43A} and at $E>150$~GeV for 3C 279 \citep{2011A&A...530A...4A}.  
However, for a falling spectrum, the flux of gamma rays with $E>200$~GeV 
($E> 150$~GeV) is dominated by the photons with energies $E\approx 
200$~GeV ($E\approx 150$~GeV).  There is no evidence of variability at higher 
energies, at which gamma rays detected from these two blazars are consistent 
with secondary gamma rays. 
For a detailed discussion of time variability in cosmic-ray-induced secondary gamma rays we refer the reader to \cite{2012arXiv1203.3787P}.

Although the exact point where the secondary signal dominates the primary is dependent on the ratio of cosmic ray luminosity to gamma ray luminosity, one can estimate the transition energy by demanding that the primary signal be attenuated by at least an order of magnitude. Since the attenuation beyond this point grows exponentially, this estimate should be fairly accurate. In Figure~\ref{fig:tau} we show the optical depth ($\tau$) for two models of EBL for two redshifts. The transition from primary to secondary photons is expected to occur between $\tau=1$ and $\tau=3$ lines.

\begin{figure}
\begin{center}
\includegraphics[width=0.95\textwidth]{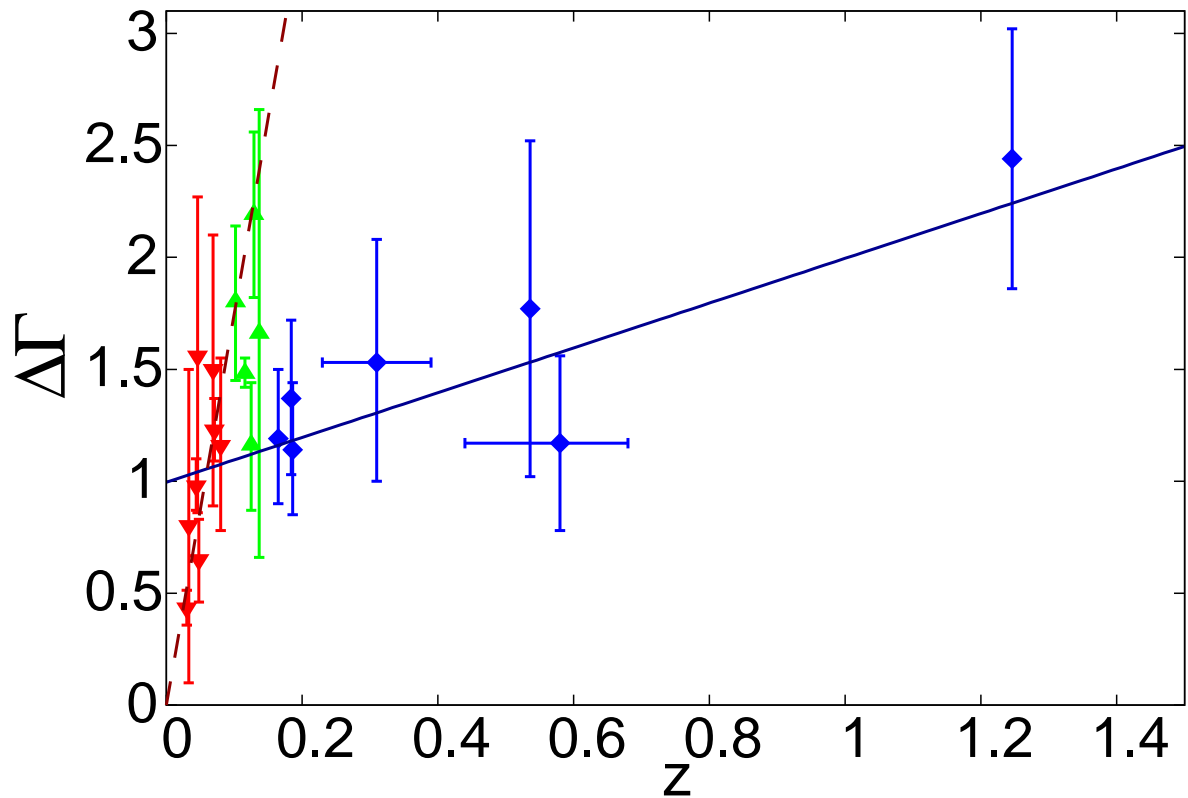}
\caption{Spectral change, $\Delta\Gamma=\Gamma_{\rm TeV}-\Gamma_{\rm GeV}$, for TeV detected blazars observed by Fermi. Data points from the Fermi Second catalog~\citep{2011arXiv1108.1420T} were separated into three sets: nearby sources (red inverted triangles), intermediate sources (green triangles) and distant sources (blue diamonds). The lines are the best fits to Equation~(\ref{eq:deltaG}) with $D=17.46$ (dashed line) and $(\Gamma_p-\Gamma_s)=0.995$ (solid line).
\label{fig:fit}}
\end{center}
\end{figure}

\begin{figure}
\begin{center}
\includegraphics[width=0.95\textwidth]{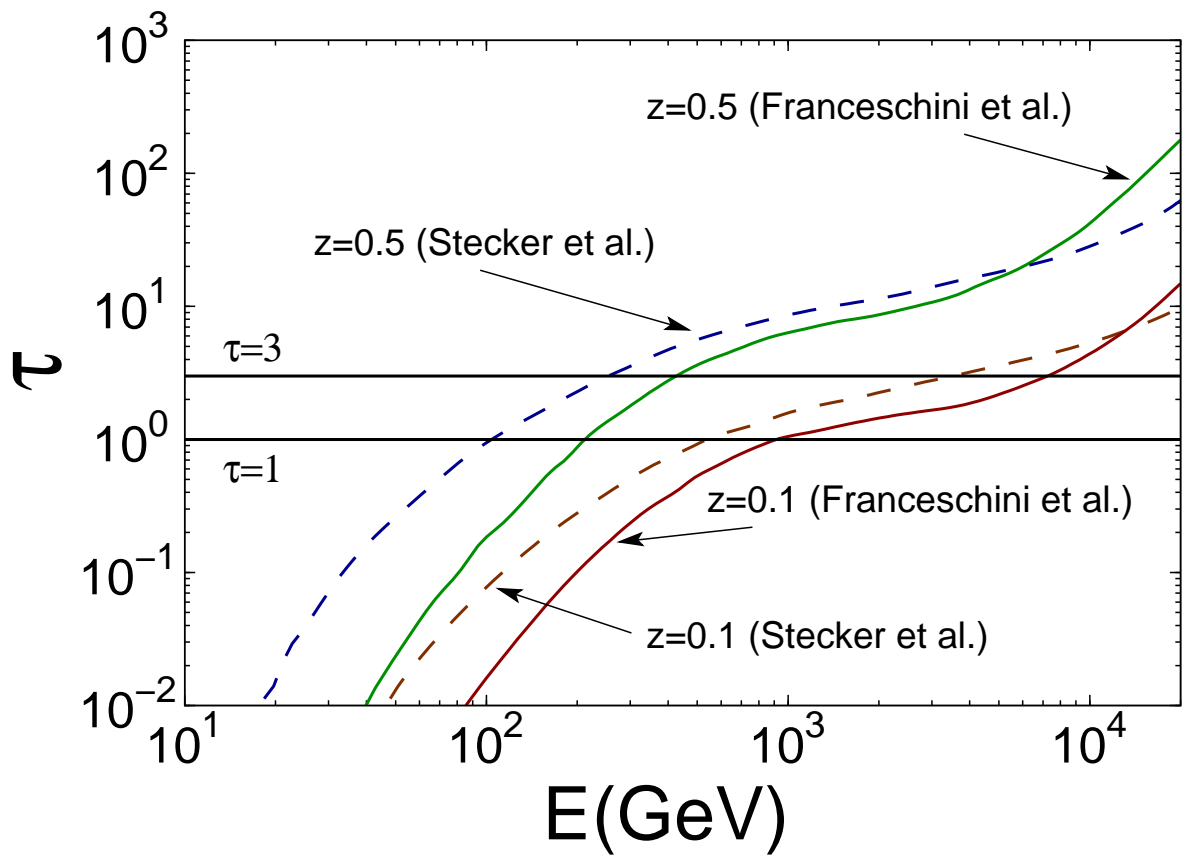}
\caption{Optical depth $\tau$ for the~\cite{2008A&A...487..837F} and \cite{2006ApJ...648..774S} models of EBL, for two redshifts. 
 The horizontal lines represent $\tau=1$ and $\tau=3$, between which the transition from primary component to secondary component takes place. 
\label{fig:tau}}
\end{center}
\end{figure}

\begin{table}
\begin{center}
\begin{tabular}{|l|l|}
\hline
Source name & Redshift \\
 \hline 
H2356-309 & 0.165 \ \mbox{\citep{1991AJ....101..821F}} \\
1ES 1218+304 & 0.184 \  \mbox{\citep{1998A&A...334..459B}} \\
1ES 1101--232&0.186 \ \mbox{\citep{1989ApJ...345..140R}} \\
S5 0716+714 & $0.31\pm0.08$ \  \mbox{\citep{2008A&A...487L..29N} } \\
3C279 & 0.536  \  \mbox{\citep{1993ApJS...87..451H}} \\
3C66A & 0.58 (0.44-0.68) \  \mbox{\citep{1993ApJS...84..109L,2010arXiv1006.4401Y,2011ApJ...726...58A}} \\
PKS 0447--439 & 1.246 \ \mbox{\citep{2012arXiv1203.4959L}} \\
 \hline
\end{tabular}
\end{center}
\caption{Distant sources and their redshifts [with references].}
\end{table}

In Figure~\ref{fig:fit} we show that the best fit for $D$ and $(\Gamma_p-\Gamma_s)$ are in good agreement with the data from the Fermi two-year catalog.  $D$ was fitted for sources with $z \lesssim 0.1$, where the primary signal is expected to dominate, and $(\Gamma_p-\Gamma_s)$ was obtained from the data for sources with $z\gtrsim0.15$, where the secondary signal is expected to dominate. The fit at high $z$ gives $\chi^2=1.05$ with 5 d.o.f. yielding the confidence probability of $P=0.96$. The agreement with the data is evident.  In particular, a recent measurement of the redshift of PKS~0447-439 \citep{2012arXiv1203.4959L}, which was detected by HESS at energies above TeV \citep{2011arXiv1105.0840Z}, is in agreement with the trend. We note that, for the relevant proton energies, $E\sim 10^{17}-10^{18}$~eV, the energy attenuation length of protons is much greater than the distance to a source at z=1.2 (see, {\em e.g.}, Figure 9 of \cite{2000PhR...327..109B}).  Therefore, the scaling laws in Eqs. (2)--(4) are valid for this extremely distant source. The inferred luminosity of this source in protons is $L_p\sim 10^{47}$ erg/s,  
assuming a $6^\circ$ ($3^\circ$ radius) beam.  This is comparable or below the  Eddington luminosity for a billion-solar-masses black hole.  
Based on the analysis of \cite{1992MNRAS.259..421C}, we estimate that several (between 1 and 10) supermassive black holes with masses $> 10^{9} M_\odot$ can be found in the $z \le 1.2$ volume with a 6-degree jet pointing at Earth.  This possibility should motivate observations of other distant blazars with atmospheric Cherenkov telescopes, as they may lead to discoveries of additional TeV sources at $z\sim 1$.

In the intermediate region, $0.1\lesssim z\lesssim 0.15$, one can detect primary signals from blazars that are brighter than average in gamma rays and accelerate fewer cosmic rays, and one can also detect secondary signals from those blazars that are more powerful cosmic-ray accelerators. Hence, in this intermediate range of redshifts, one can expect both spectral slopes to be present.  This is, indeed, evident from the data plotted in Figure~\ref{fig:fit}, where the blazars with $0.1 \lesssim z \lesssim 0.15$ have a broader spread of spectral indices, and some of the blazars tend to the {\em primary} curve, while other blazars agree with the  
{\em secondary} scaling law.  

Secondary gamma rays can contribute to point sources only if intergalactic magnetic fields (IGMF) are in the range 0.01~fG~$ < B < 30$~fG~\citep{2011APh....35..135E}, although these bounds can be affected by the source variability~\citep{2011ApJ...733L..21D,2011ApJ...727L...4D}.  
The lower and the upper limits were obtained by \cite{2011APh....35..135E} for the case of line-of-sight interactions using only the spectral data, 
with no reference to the source morphology. The agreement of spectral evolution with the data strengthens these inferences regarding IGMFs. 
In the upper part of this range, angular resolution of {\em Fermi} should be good enough to resolve halos of AGN images~\citep{1994ApJ...423L...5A}, which can provide an independent measurement.

\section{Conclusions}

We have generalized the spectral evolution relation of \cite{2006ApJ...652L...9S,2010ApJ...709L.124S} to include the contribution of cosmic ray interactions along the line of sight. The predicted scaling with redshift agrees with the data, which lends further support to the hypothesis of cosmic ray acceleration in blazars and to the inferences regarding universal backgrounds and AGN powers made by \cite{2010APh....33...81E,2010PhRvL.104n1102E,2011ApJ...731...51E,2011arXiv1107.5576M,2011APh....35..135E,2012ApJ...745..196R,2012arXiv1203.3787P}.

We thank F.~Aharonian, J.~Beacom, C.~Dermer, O.~Kalashev, S.~Razzaque, and F.~Stecker for helpful comments and discussions. This work was supported in part by DOE grant DE-FG03-91ER40662.

  

\end{document}